\pdfoutput=1
\documentclass{aa}
\usepackage{graphicx}
\usepackage{amsmath}
\usepackage{booktabs}
\usepackage{microtype}
\usepackage[
colorlinks=true,
linkcolor=blue,
urlcolor=blue,
citecolor=blue,
anchorcolor=blue]
{hyperref}
\usepackage[varg]{txfonts}
\usepackage{natbib}
\usepackage[modulo,switch]{lineno}

\newcommand{\eq}[1]{
\begin{equation}
#1
\end{equation}}
\newcommand{\sci}[2]{#1\times10^{#2}}
\newcommand{\st}[1]{_\text{#1}}
\newcommand{\bracfrac}[2]{\left(\frac{#1}{#2}\right)}
\newcommand{\K}{\rm\thinspace K}
\newcommand{\cm}{\rm\thinspace cm}
\newcommand{\s}{\rm\thinspace s}
\newcommand{\cmpssq}{\hbox{$\cm\s^{-2}\,$}}
\newcommand{\uHz}{\hbox{\rm\thinspace $\mu$Hz}}

\begin{document}

\title{MESA meets MURaM}%
\subtitle{Surface effects in main-sequence solar-like oscillators
  computed using three-dimensional radiation hydrodynamics
  simulations}%
\titlerunning{MESA meets MURaM}%

\author{W.~H.~Ball\inst{1,2} 
  \and B.~Beeck\inst{2}
  \and R.~H.~Cameron\inst{2}
  \and L.~Gizon\inst{2,1}}%
\institute{
  Institut f\"ur Astrophysik, Georg-August-Universit\"at G\"ottingen, 
  Friedrich-Hund-Platz 1, 37077 G\"ottingen, Germany\\
  \email{wball@astro.physik.uni-goettingen.de}\and 
  Max-Planck-Institut f\"ur Sonnensystemforschung, 
  Justus-von-Liebig-Weg 3, 37077 G\"ottingen, Germany}

\abstract{Space-based observations of solar-like oscillators have
  identified large numbers of stars in which many individual mode
  frequencies can be precisely measured.  However, current stellar
  models predict oscillation frequencies that are systematically
  affected by simplified modelling of the near-surface layers.}
{We use three-dimensional radiation hydrodynamics simulations to
  better model the near-surface equilibrium structure of dwarfs with
  spectral types F3, G2, K0 and K5, and examine the differences
  between oscillation mode frequencies computed in stellar models with
  and without the improved near-surface equilibrium structure.}%
{We precisely match stellar models to the simulations' gravities and
  effective temperatures at the surface, and to the temporally- and
  horizontally-averaged densities and pressures at their deepest
  points.  We then replace the near-surface structure with that of the
  averaged simulation and compute the change in the oscillation mode
  frequencies.  We also fit the differences using several parametric
  models currently available in the literature.}%
{The surface effect in the stars of solar-type and later is
  qualitatively similar and changes steadily with decreasing effective
  temperature.  In particular, the point of greatest frequency
  difference decreases slightly as a fraction of the acoustic cut-off
  frequency and the overall scale of the surface effect decreases.
  The surface effect in the hot, F3-type star follows the same trend
  in scale (i.e. it is larger in magnitude) but shows a different
  overall variation with mode frequency.  We find that the two-term
  fit by Ball \& Gizon (2014) is best able to reproduce the surface
  terms across all four spectral types, although the scaled solar term
  and a modified Lorentzian function also match the three cooler
  simulations reasonably well.}
{Three-dimensional radiation hydrodynamics simulations of near-surface
  convection can be averaged and combined with stellar structure
  models to better predict oscillation mode frequencies in solar-like
  oscillators.  Our simplified results suggest that the surface effect
  is generally larger in hotter stars (and correspondingly smaller in
  cooler stars) and of similar shape in stars of solar type and
  cooler.  However, we cannot presently predict whether this will
  remain so when other components of the surface effect are included.}


\maketitle

\section{Introduction}

The modern era of asteroseismology, driven principally by space-based
missions like CoRoT \citep{corot} and \textit{Kepler} \citep{kepler},
is providing a wealth of data for many oscillating stars of various
types.  Among these is a large number of solar-like oscillators in
which individual mode frequencies can be precisely measured.  When
combined with non-seismic observations, these frequencies have the
potential to tightly constrain both the parameters and physics of
stellar models.  

\begin{figure}
\includegraphics[width=85mm]{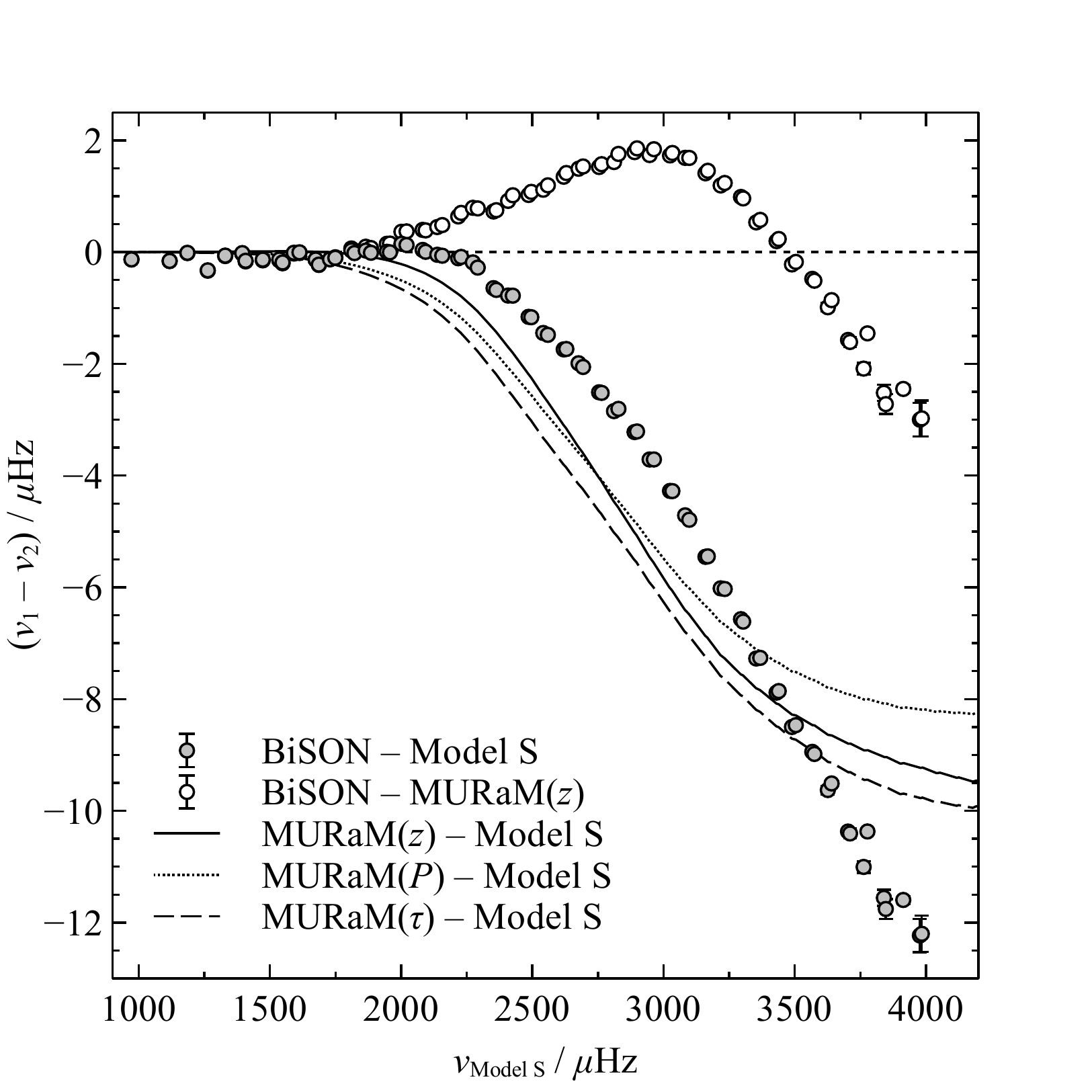}
\caption{Frequency differences between BiSON observations of
  low-degree solar oscillations and Model S before (grey points) and
  after (white points) the near-surface layers are replaced by the
  profile of the G2-type MURaM simulation averaged over time and
  surfaces of constant geometric depth. The solid line shows the
  difference between the frequencies before and after replacing the
  near-surface layers.  In addition, the dotted and dashed lines show
  the frequency differences when the MURaM simulation is instead
  averaged over surfaces of constant pressure $P$ or optical depth
  $\tau$.  With the averaged MURaM profile, the residual frequency
  difference is reduced to at most about $3\uHz$.  The frequency
  changes are sensitive to the averaging method at about the $1\uHz$
  level, which is smaller than the surface effect itself.}
\label{f:modelS}
\end{figure}

A major obstruction to exploiting these data is the systematic
difference between modelled and measured frequencies, known from
standard solar models and observations of solar oscillations, of which
an example is shown in Fig.~\ref{f:modelS}.  The grey points show the
differences between mode frequencies computed for a standard solar
model \citep[Model S,][]{modelS} and observations of low-degree modes
($\ell\leq3$) from the Birmingham Solar Oscillation Network
\citep[BiSON `quiet sun' frequencies,][]{broomhall2009,davies2014}.
The differences broadly increase with frequency and appear to be
independent of angular degree, from which it is inferred that the
cause is located near the surface of the star, and is known as the
\emph{surface effect} or \emph{surface term}.  The behaviour of the
surface effect changes as the frequencies approach the acoustic
cut-off frequency\footnote{Here, we define the acoustic cut-off
  frequency to be the maximum of $c_s/4\pi H_P$, where $c_s$ is the
  sound speed and $H_P$ is the pressure scale height.} $\nu\st{ac}$
and becomes increasingly sensitive to the structure of the upper
atmosphere (i.e. above the photosphere), which is also uncertain but
not studied in detail here.  In the subsequent discussion, it should
be noted that part of the surface effect is presumably addressed by
improved atmospheric structure.

The surface effect has several causes, which can broadly be classified
as flaws in the equilibrium structure of the star (\emph{model}
physics) and flaws in the computation of the oscillation frequencies
(\emph{modal} physics).  On the structure side, it is well-known that
mixing-length theories of convection incorrectly predict stellar
structure near the surface, where convection is inefficient, the
temperature gradient is strongly superadiabatic, vertical flows are
asymmetric and rms convective velocities reach up to 15 per cent of
the local sound speed.  These effects have already been considered for
the Sun.  \citet{schlattl1997} calibrated a solar model with an
independent atmosphere model down to optical depth $\tau\approx20$ and
a variable mixing-length parameter calibrated to two-dimensional
hydrodynamic simulations of the near-surface layers, and obtained
better agreement between observations and their modelled mode
frequencies.  \citet{rosenthal1999} studied frequency differences in
high-degree modes, confined to the convective envelope, between
mixing-length models and detailed three-dimensional hydrodynamic
simulations.  \citet{piau2014} patched similar simulations onto
complete solar models, like we do here, but they only studied the
effect on solar mode frequencies.

On the oscillation side, frequencies are typically computed in the
adiabatic approximation and neglect the perturbation to the turbulent
pressure (which is usually absent from the stellar model anyway).  In
addition, the oscillation calculations ignore the modifications to
wave propagation caused both by the small-scale random motions of the
fluid and the large-scale motions: slow, wide upflows and fast, narrow
downflows.  \citet{brown1984} demonstrated that even in a symmetric
flow, a wave's mean phase speed is retarded, though the effect on mode
frequencies is stronger at higher angular degree $\ell$.
\citet{murawski1993a,murawski1993b} studied how the f-mode in
particular is affect by multiple scattering through a random flow and
were able to partly reconcile the deviation of the mode frequencies
from the simple dispersion relation $\omega^2=gk$, where $\omega$ is
the (angular) mode frequency, $g$ the surface gravity, and $k$ the
wavenumber.  \citet{jishnu2015} used the method of homogenization
\citep[see also][]{hanasoge2013} to develop a formalism for frequency
shifts in the limit of modes with horizontal wavelengths much longer
than the scale of granulation: true for low-degree modes.  However, a
satisfactory formalism for the interaction of solar waves with
time-dependent turbulence is still lacking.  In addition, a small
component of the surface effect varies with the solar magnetic
activity cycle and is therefore usually associated with structural
changes caused by variations in the Sun's magnetic field
\citep{libbrecht1990, goldreich1991}.

Recently, multiple research groups \citep[e.g.][]{beeck2013a,
  trampedach2013, ludwig2009} have computed three-dimensional
radiation hydrodynamics simulations of the convective near-surface
layers of stars with various surface gravities and effective
temperatures.  Unlike standard stellar models, these simulations model
convection from first principles and more accurately describe the
highly superadiabatic near-surface layers and atmospheres of stars.
Thus, the simulations have the potential to improve our predictions of
the stars' mode frequencies, because they better model the layer in
which simplified modelling causes the surface effect.  Already,
\citet{trampedach2014a} have used their simulations to calibrate
atmospheric $T(\tau)$ relations, which will partly mitigate the
surface effect.

\citet{beeck2013a} modelled surface convection in dwarfs of six
spectral types, ranging from F3 to M2.  We experimented with simply
replacing the near-surface layers of Model S with the horizontally-
and temporally-averaged structure of their G2 simulation and computing
the oscillation mode frequencies using this \emph{patched} model.  The
solid curve in Fig.~\ref{f:modelS} shows the frequency shift induced
by replacing the near-surface layers with the simulation data averaged
over time and at constant geometric depth, and the white points show
the residual difference between the observations and the model
frequencies.  The patched model is not perfect but the magnitude of
the surface effect is reduced from over about $12\uHz$ to at
  most about $3\uHz$, and our calculation compares well with previous
results \citep[e.g. Fig.~7 of][]{piau2014}.

Though the MURaM simulation and Model S have not been calibrated to
each other, both match the same deeper, near-adiabatic structure of
the Sun.  In solar-calibrated stellar models like Model S, the choice
of mixing-length parameter fixes the entropy jump across the
near-surface superadiabatic layer and, consequently, also fixes the
adiabat of the deep convection zone \citep{gough1976}.  The MURaM
simulation is sufficiently realistic that, given the Sun's effective
temperature, surface gravity and composition, the same adiabat is
recovered.

The dotted and dashed curves in Fig.~\ref{f:modelS} show the frequency
changes induced by instead replacing the near-surface layers with the
simulation data averaged at constant pressure or optical depth.  The
variation in the surface term shows that there is some uncertainty
induced by the averaging process, but this uncertainty is smaller than
the overall scale of the frequency changes.  For the rest of the
paper, all averages of simulation properties were taken at constant
geometric depth.

\begin{figure}
\includegraphics[width=85mm]{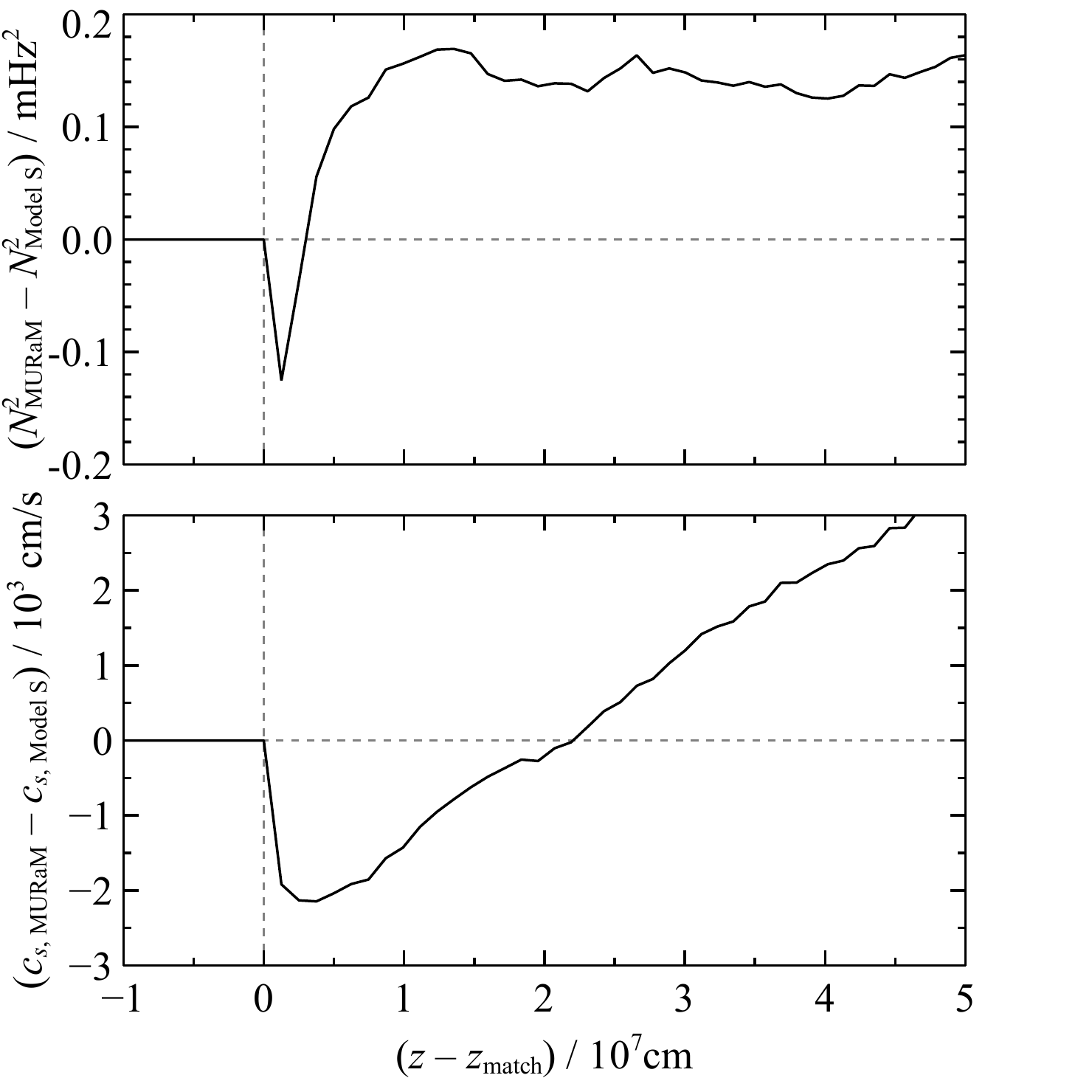}
\caption{Structural differences between the patched and
    unpatched models near the matching point.  The upper panel shows
    differences in the Brunt--V\"ais\"al\"a frequency while the lower
    panel shows differences in the sound speed.}
\label{f:match}
\end{figure}

Fig.~\ref{f:match} shows the differences in the Brunt--V\"ais\"al\"a
frequency $N^2$ and sound speed $c_s$ at the matching point.  Though
the change in the Brunt--V\"ais\"al\"a is fractionally large, the
matching point is inside the convection zone, where $N^2$ is negative.
The mode frequencies are large and apparently unaffected, though this
would not be the case for non-radial modes in more evolved stars.  The
fractional sound speed difference is less than $0.2$ per cent.

The change in the frequencies caused only by the modification of the
equilibrium stellar structure is just one of the many causes of the
surface effect.  As noted in the extensive discussion by
\citet{rosenthal1997}, it is difficult to predict how these many
effects will combine to give precisely the correct surface term, but
this does not mean that it is not worth studying the individual
effects in isolation.  Thus, the results here should be interpreted
with the caveat that they are only one component of the surface effect
and have been computed in a simplified fashion.

Motivated by the positive result for the Sun, we computed stellar
models to match the properties of four of the simulations by
\citet{beeck2013a} in the same way, listed in Table~\ref{t:data}, and
computed the change in mode frequencies induced by replacing the
surface layers with the averaged simulation data.  We present here a
preliminary exploration of how the surface effect varies with spectral
type on the main sequence, based on these simulations.  This is the
same procedure as followed in the recent results by \citet{sonoi2015}.
Their models A and B have similar atmospheric parameters to our G2 and
F3 simulations.  Their other models extend to more evolved stars,
whereas ours extend to cooler main-sequence stars.  In
Section~\ref{s:methods}, we briefly describe the simulations, the
stellar models and how we matched them.  In Section~\ref{s:results},
we present the frequency shifts and compare them with several
parametric fits in the literature, before concluding in
Section~\ref{s:conclusion}.

\begin{table*}
  \caption{Parameters of the MURaM simulations and MESA models along with
    best-fitting parameters for analytic corrections available in the literature.
    The coefficient $a_3$ corresponds to the one-term \emph{cubic} fit by \citet[][eq.~\ref{e:bg1}]{ball2014},
    and the coefficients $b_{-1}$ and $b_3$ to their two-term \emph{combined} fit (eq.~\ref{e:bg2}).  Parameters
    $p_0$ and $p_1$ are the amplitude and index of a power-law 
    correction (eq.~\ref{e:powerlaw}), as proposed by \citet{kbcd2008}, 
    and $s_0$ and $s_1$ are the amplitude and index of the modified Lorentzian
    (eq.~\ref{e:sonoi}) suggested by \citet{sonoi2015}.
    The last parameter $c$ corresponds to the amplitude of a scaled solar surface term, computed
    as in \citet{schmitt2015}.  The last row of the table gives the values found by calibrating
    the surface term for Model S with respect to BiSON observations.}
\begin{center}
\begin{tabular}{ccccccccccccc}
\toprule
& $\log g$ & $T\st{eff}$ & $\alpha\st{MLT}$ & $\nu\st{ac}$ &
$a_3$ & $b_{-1}$ & $b_{3}$ & $p_0$ & $p_1$ & 
$s_0$ & $s_1$ & $c$ \\
Units & $\cmpssq$ & $\K$ & & $\uHz$ & 
$10^{-7}\uHz$ & $10^{-8}\uHz$ & $10^{-7}\uHz$ & $\uHz$ & & 
$10^{-3}$ & & \\
\midrule
F3 & $4.301$ & $6893.2\pm6.4$ & $1.63$ & $3\,283$ &
$-4.05$ & $-3.97$ & $-2.67$ & $-44.0$ & $1.88$ & 
$-19.69$ & $3.51$ & $3.26$ \\

G2 & $4.438$ & $5764.4\pm7.4$ & $1.74$ & $5\,022$ & 
$-2.55$ & $-3.33$ & $-1.71$ & $-20.0$ & $1.96$ & 
$-5.68$ & $5.25$ & $1.44$ \\

K0 & $4.609$ & $4855.6\pm5.5$ & $1.85$ & $8\,113$ & 
$-1.65$ & $-2.85$ & $-0.99$ & $-13.2$ & $1.87$ & 
$-2.37$ & $5.58$ & $0.76$ \\

K5 & $4.699$ & $4367.9\pm2.0$ & $1.44$ & $10\,367$ &
$-1.77$ & $-1.90$ & $-1.29$ & $-12.1$ & $2.37$ & 
$-1.40$ & $7.32$ & $0.41$ \\

\midrule
Model S & & & & $5\,212$ & $-1.92$ & $0.29$ & $-2.15$ & $-71.8$ & $5.37$ &
$-3.59$ & $11.26$ & \\
\bottomrule
\end{tabular}
\end{center}
\label{t:data}
\end{table*}

\section{Methods}
\label{s:methods}

\subsection{MURaM models}
\label{s:muram}

The simulations of the near-surface convection used in this work were
computed by \citet{beeck2013a}, to which the reader is referred for
additional details and references about the simulations and the code
used to produce them, MURaM.  In short, MURaM is a three-dimensional,
time-dependent, compressible radiative magnetohydrodynamics (MHD) code,
developed jointly by groups at the University of Chicago and the Max
Planck Institute for Solar System Research \citep{voegler2005}.  The
non-grey radiation transport scheme is based on the method of short
characteristics \citep{kunasz1988} and uses opacity binning
\citep{nordlund1982} based on opacity distribution functions from
ATLAS9 \citep{kurucz1993}.  The code uses the OPAL equation of state
\citep{rogers1996} with the solar composition of \citet{anders1989}.

The three-dimensional, time-dependent simulations were averaged over
time and at constant geometric depth $z$ (where $z=0$ corresponds to
the average depth of the $\tau\approx1$ surface) to produce
one-dimensional profiles of density, pressure and sound speed as
functions of depth.  This is a simplification: it is not known how to
correctly average the fluid such that the average oscillations are
reproduced.
The averaged simulation profiles were then used to replace the
near-surface layers of the stellar models, described below.
Specifically, we used profiles for the averaged pressure, density (and
their gradients, computed from finite differences of their logarithms)
and sound speed as functions of depth.  This implies that although the
MURaM simulations at a given point and time satisfy the same equation
of state as the stellar models, the averages do not.  The adiabatic
index was computed from the sound speed, so that although this
adiabatic index is not the same as the simulation average, it
reproduces the average sound speed for the oscillation calculation.
Finally, the Brunt-V\"ais\"al\"a frequency was computed from the
adiabatic index, pressure gradient and density gradient.  Like
\citet{sonoi2015}, we assumed that the relative Lagrangian
perturbation to the turbulent pressure is the same as that of the gas
pressure, which \citet{rosenthal1999} referred to as the \emph{gas
  $\Gamma_1$ approximation}.

\subsection{MESA models}
\label{s:mesa}

Stellar models were computed using the Modules for Experiments in
Stellar Astrophysics
\citep[MESA\footnote{\url{http://mesa.sourceforge.net/}}, revision
7624,][]{paxton2011,paxton2013,paxton2015}.  As far as was possible,
we used default options for the stellar models.  Each model was
initialized on the pre-main-sequence with uniform composition and
central temperature $\sci{9}{5}\K$ and evolved until either hydrogen
was exhausted at the centre or $\chi^2$ had increased to more than 100
times the minimum value for that run (see below for how $\chi^2$ was
defined).  The overall metallicity (i.e. $Z/X$) and metal mixture of
the stellar models was chosen to be that of \citet{anders1989} to
match the composition of the MURaM simulations.  We used a helium
abundance of $0.27431$.  To ensure that the surface composition
matched the simulations, atomic diffusion and extra mixing were not
included.  Opacities were taken from the OPAL tables
\citep{iglesias1996} and \citet{ferguson2005} at high and low
temperatures, respectively.  The opacity tables were computed with the
nearest available solar mixture, that of \citet{grevesse1993}.  The
equation of state tables are based on the 2005 update to the OPAL
tables \citep{rogers2002}.  Nuclear reaction rates are taken from the
NACRE compilation \citep{angulo1999} or, if not available there, from
\citet{caughlan1988}.  Convection was described using mixing-length
theory \citep{boehm-vitense1958}.  We used an Eddington grey
atmosphere at the surface boundary, integrated to an optical depth of
$\tau=10^{-4}$.

\subsection{Fitting method}
\label{s:fits}

For each simulation, we first matched stellar models with six mixing
length parameters $\alpha$ to the simulation's $T\st{eff}$ and $\log
g$ using the downhill simplex \citep{nelder1965} optimization
implemented in MESA.  We optimized the masses and ages for models with
fixed mixing-length parameters $\alpha$ between $1.5$ and $2.0$ in
steps of $0.1$.  The objective function was a standard $\chi^2$, with
the uncertainty on $T\st{eff}$ taken as the rms variation of the
temperature listed by \citet{beeck2013a}, and the uncertainty in $\log
g$ set to $0.001\,\mathrm{dex}$.  Because we optimized two parameters (mass and age)
for two observations ($T\st{eff}$ and $\log g$), the choice of
uncertainties is in essence arbitrary.

We then patched the averaged MURaM profiles onto the stellar models by
finding the depth at which the pressure of the stellar model matched
the pressure at the base of the MURaM profile, and replacing the
stellar model data with the averaged data from the simulation.  We
inspected the difference between the densities at the matching point,
and computed additional stellar models with intermediate mixing-length
parameters until a sufficiently accurate match was found.  As a
cross-check, we inspected the sound speed profile to ensure that it
too was sufficiently smooth.  The best-fitting values of $\alpha$ are
also listed in Table~\ref{t:data}.  With this procedure, we calibrated
$\alpha$ to a precision of about $0.01$.

The best-fitting model of type K5 has a lower mixing-length parameter
$\alpha$ than the other simulations.  This goes against trends
determined by matching 1D mixing-length theory atmospheres to 2D and
3D simulations \citep[e.g.][]{ludwig1999, trampedach2014b}.  Though we
have not found an obvious explanation for this, we note that the
convective envelope in the stellar model extends above the
photosphere, but the photosphere boundary condition assumes that the
luminosity is transported only by radiation.  In addition, this
stellar model is so young that it still has a residual convective core
from its pre-main-sequence contraction.  Regardless of the
unexpectedly low mixing-length parameter, we confirmed that the
structure of the K5 stellar model matches the simulation profile in
its deepest layers, just like for the other three stellar types.

The process above gave us two models for each spectral type: the
original stellar model and the patched model, where the near-surface
layers were replaced by the averaged MURaM simulations.  Both models
share the same internal structure below the matching point and
therefore have the same luminosity, but the patched models have
slightly larger radii and lower surface temperatures.  We computed the
adiabatic oscillation frequencies for both models using ADIPLS
\citep{adipls} and took the frequency shift as the difference between
the frequencies for modes found for both models.  The process is
similar in spirit to the study by \citet{piau2014} but we have neither
incorporated information from the MURaM simulations into the stars'
evolution nor considered the effects of magnetic fields.  Our method
is in essence identical to the method of \citet{sonoi2015} though we
have used low-degree modes up to $\ell=3$, whereas they restricted
themselves to radial modes.

We close by noting that it is possible to compute the frequency shifts
from structure kernels by treating the differences between the patched
models and the original stellar models as a structural perturbation.
Given that the fractional structure differences are large,
quantitative results are questionable.  Nevertheless, we performed
this calculation for Model S, treating the differences as sound speed
and density perturbations, and obtained qualitatively similar results,
dominated by the density perturbation.

\begin{figure}
\includegraphics[width=85mm]{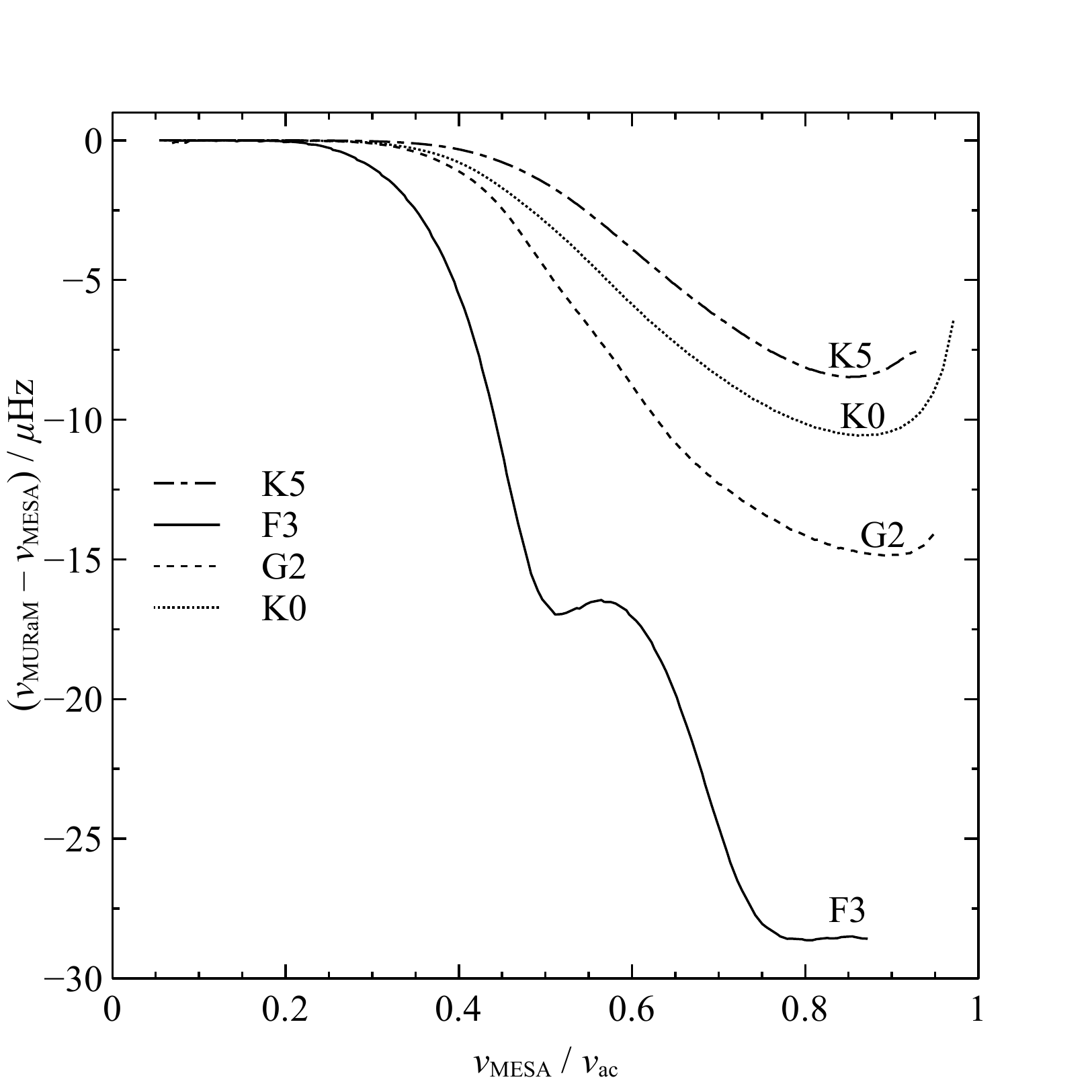}
\caption{Frequency differences between the stellar models and patch
  models computed for all four simulations, with the horizontal axis
  rescaled by the acoustic cut-off frequency $\nu\st{ac}$.  The shapes
  of the frequency differences as a function of frequency are similar
  for the three cooler simulations (G2, K0 and K5), whereas the
  difference for the F3 simulation is more complicated.}
\label{f:all}
\end{figure}

\begin{figure}
\includegraphics[width=85mm]{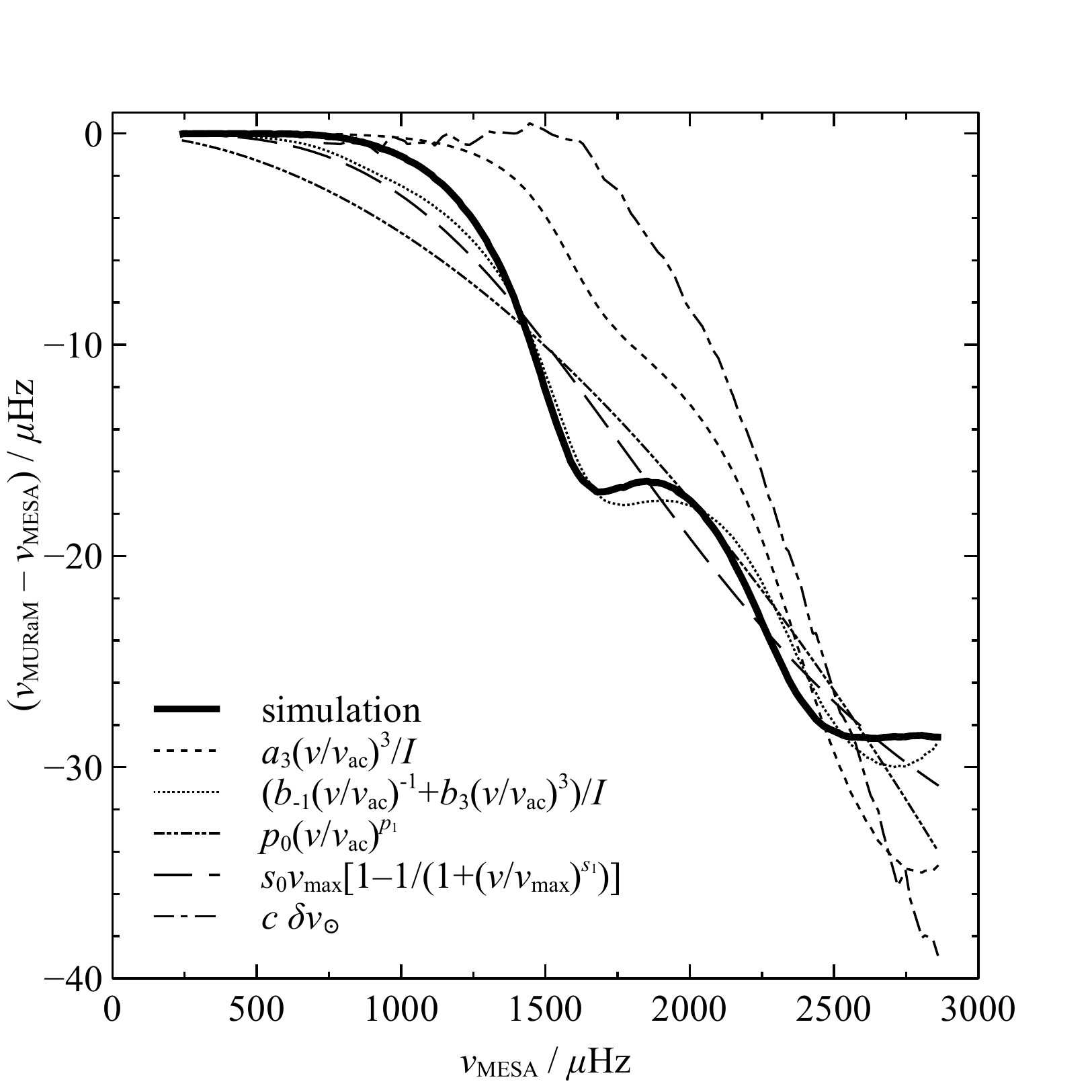}
\caption{Frequency differences between a stellar model calibrated to
  the F3 simulation, and the patched model, in which the near-surface
  layers were replaced by the averaged simulation profile.  The shaded
  region indicates the uncertainty induced by the uncertainty of about
  $0.01$ in the mixing-length parameter.  Additional curves show the
  parametric fits of the one-term correction by \citet{ball2014}
  (dashed), their two-term fit (dotted), a power law (dot-dot-dashed),
  the modified Lorentzian of \citet{sonoi2015} (long-dashed), and a
  scaled solar correction (dot-dashed).  The two-term fit is clearly
  better able to capture the distinct shape of the frequency
  difference.}
\label{f:F3}
\end{figure}

\begin{figure}
\includegraphics[width=85mm]{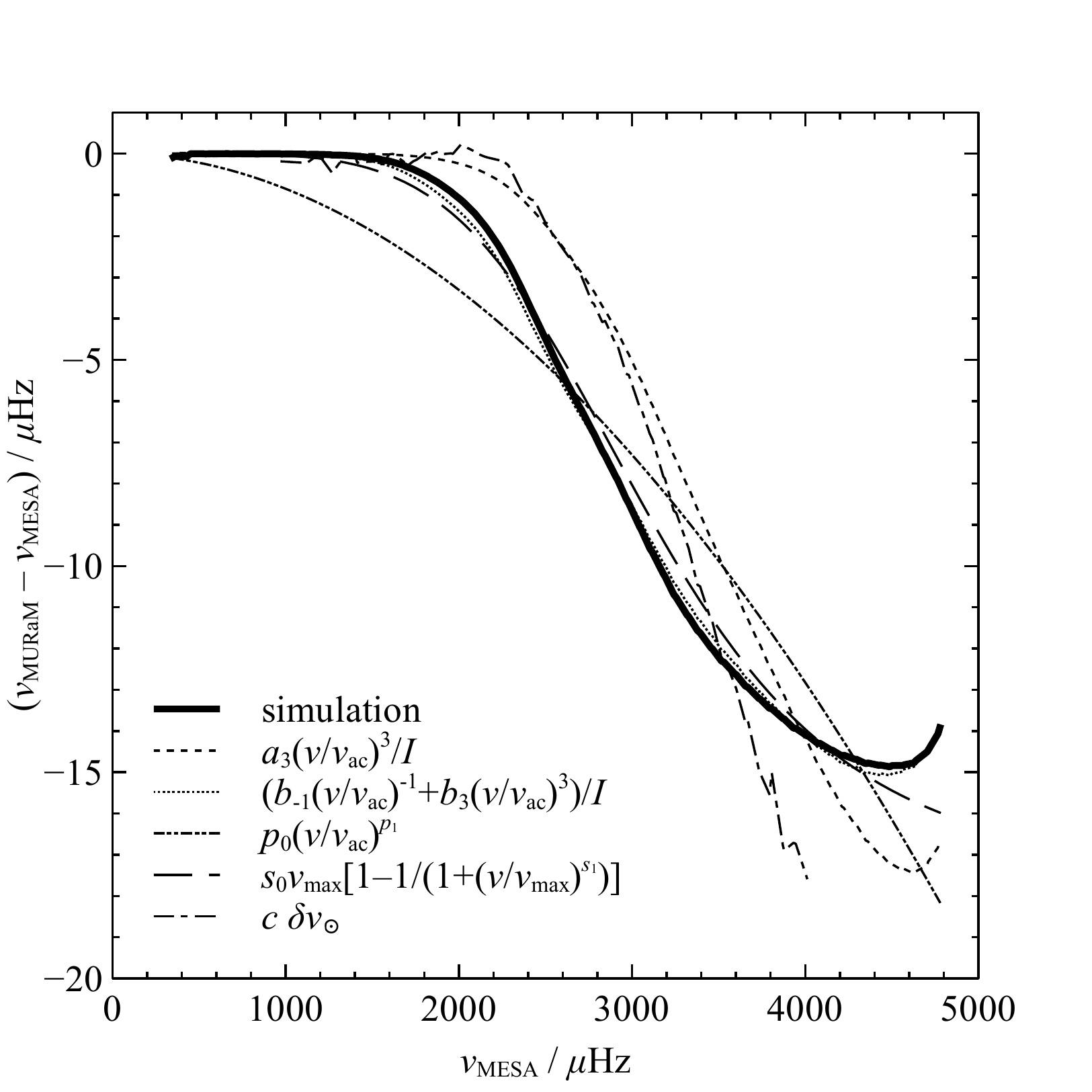}
\caption{As in Fig.~\ref{f:F3}, but for the G2 star.  The two-term fit
  is nearly perfect across the whole frequency range.}
\label{f:G2}
\end{figure}

\begin{figure}
\includegraphics[width=85mm]{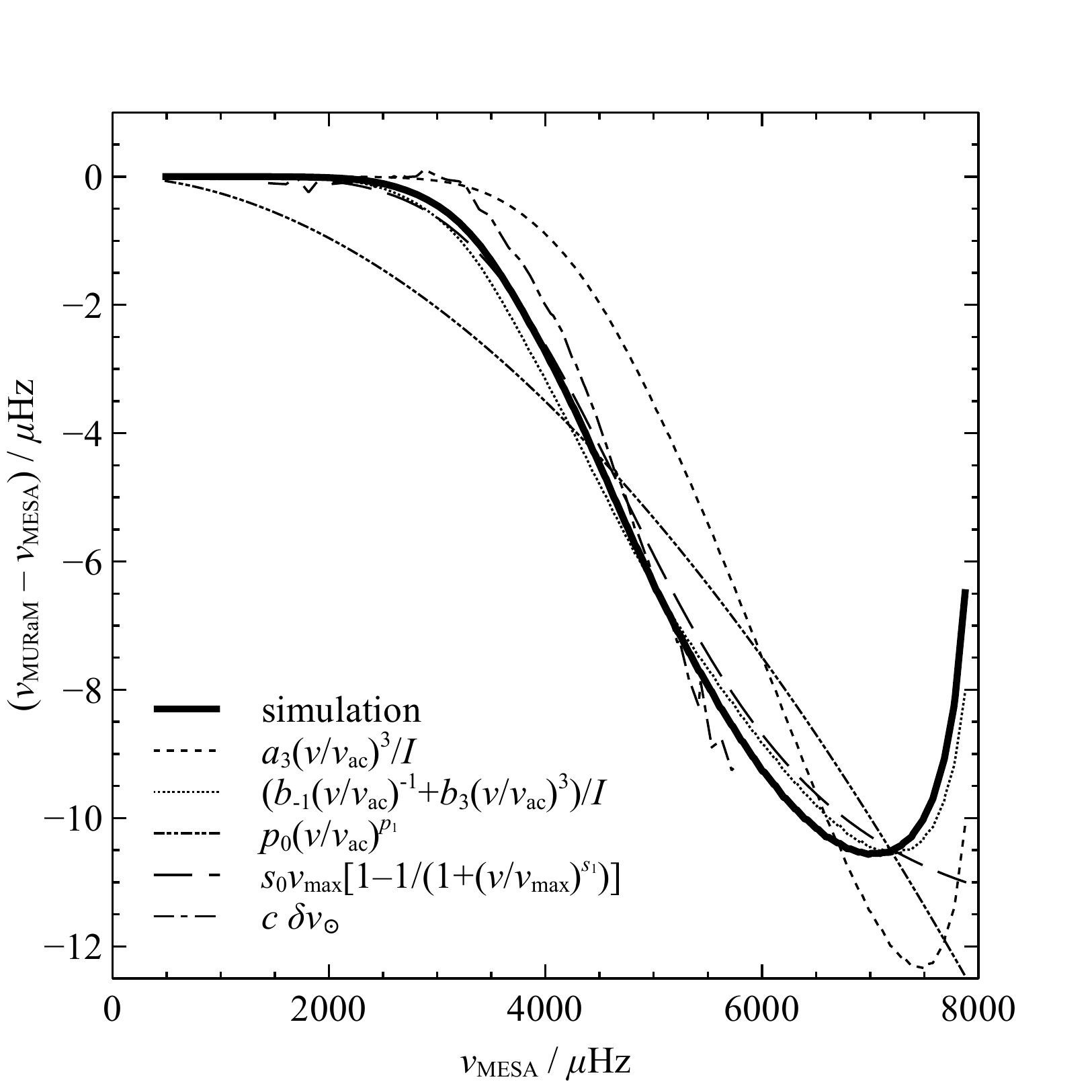}
\caption{As in Fig.~\ref{f:F3}, but for the K0 star.  The two-term fit
  again fits best, but the modified Lorentzian and the scaled
  solar correction also reproduce the shift reasonably well.}
\label{f:K0}
\end{figure}

\begin{figure}
\includegraphics[width=85mm]{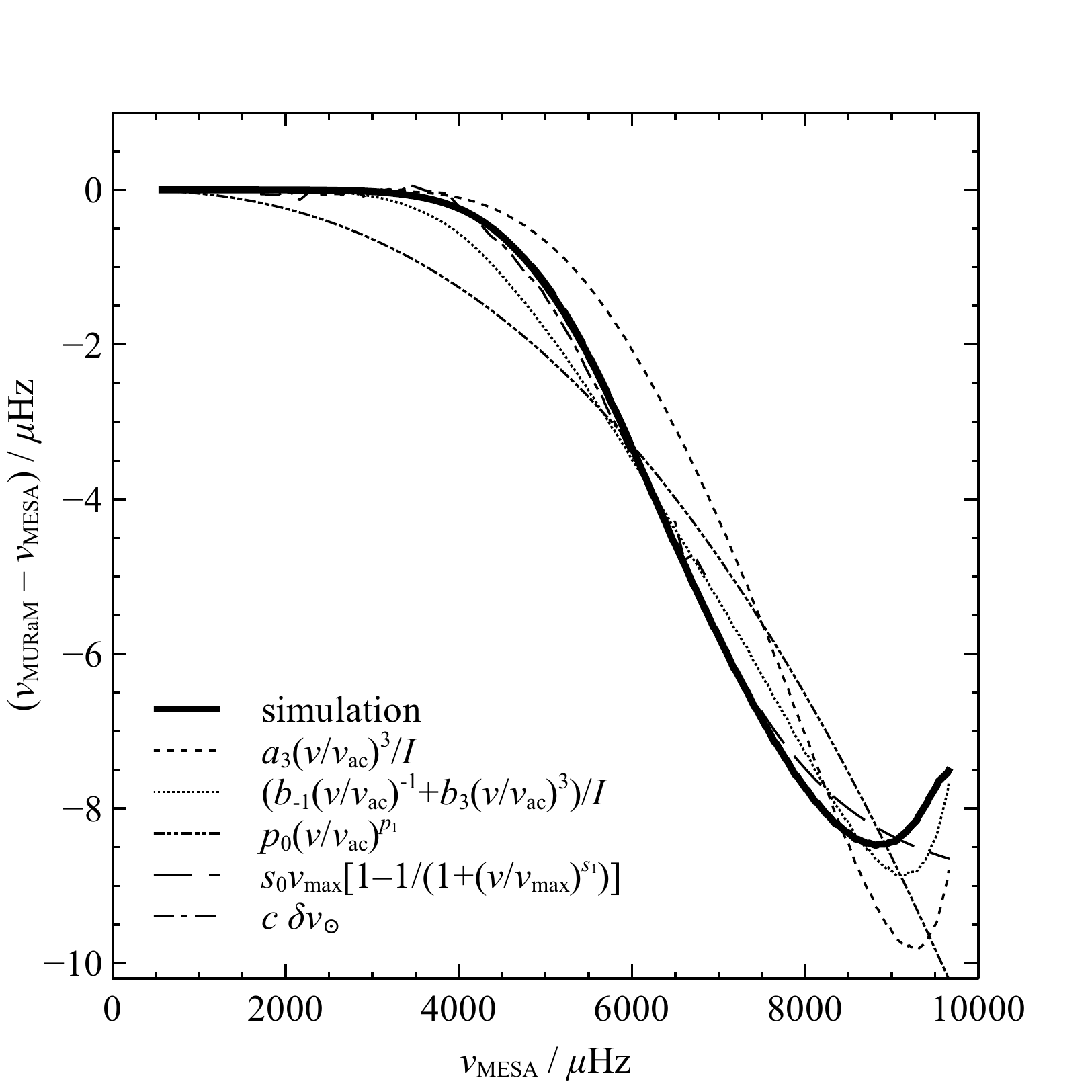}
\caption{As in Fig.~\ref{f:F3}, but for the K5 star.  As for the K0
  star, the two-term fit, modified Lorentzian and scaled solar
  correction match quite well.}
\label{f:K5}
\end{figure}

\section{Results}
\label{s:results}

\subsection{Overall features}

Fig.~\ref{f:all} shows the frequency differences as a function of
frequency for all four simulations, with the frequencies divided by
the acoustic cut-off frequency $\nu\st{ac}$.  All the surface effects
are negative in sign at high frequencies, as for the Sun, in the sense
that stellar models overpredict the mode frequencies.  In addition,
the overall scale of the surface effect decreases with decreasing
effective temperature.  There is also a clear change around
$0.3\,\nu\st{ac}$, below which the surface effect is smaller than
about $1\uHz$ across all the stars.  \citet{jcd1997} showed that this
is caused by the near-cancellation of different contributions to
Eulerian surface perturbations.  Lagrangian perturbations are confined
closer to the surface, above the upper turning point of the
low-frequency modes.

The surface effects in the three cooler spectral types (G2, K0 and K5)
are similar in shape.  The largest absolute differences occur at
decreasing relative frequency because of the shape of the mode inertia
curve.  That is, as the temperatures decrease, the minimum mode
inertia occurs at slightly lower relative frequency.  As expected, the
results depend on the choice of underlying stellar model, since the
surface shift in the calibrated G2 model differs by up to 5 $\uHz$
compared to the frequency shift computed for Model S.

The shape of the surface effect in the F3 model is notably different,
showing multiple bends as a function of frequency.  This is mostly
because the mode inertiae are a more complicated function of
frequency.  The bend in the surface effect around $-17\uHz$
corresponds to the first minimum of the mode inertiae, and the
additional bend around $-30\uHz$ corresponds similarly to a second
minimum.  The patched model also shows a jump in $\Gamma_1$ of about
$0.02$ at the matching point because helium is partially ionized at
the base of the MURaM simulation.  Such abrupt changes in the stellar
structure, known as \emph{acoustic glitches}
\citep[e.g.~][]{houdek2007}, create features in the frequencies that
oscillate as a function of frequency.  In this case, however, it
appears that the shape of the mode inertia curve dominates the change
in the frequencies. This is made clearer by the parametric fits in
Section~\ref{ss:pfits}.

Though already widely accepted, the shapes of the frequency
differences confirm that most of the modes observed in main-sequence
solar-like oscillators of all spectral types are affected by surface
effects.  The frequency at which maximum oscillation power is
observed, known as $\nu\st{max}$, is roughly $0.6\,\nu\st{ac}$, with a
typical FWHM of about half of $\nu\st{max}$
\citep[i.e. $0.3\,\nu\st{ac}$,][]{chaplin2011}.  Hence, our results
confirm that one should expect surface effects to affect all modes
observed within one FWHM of the power envelope.  This motivates
observations using line-of-sight velocities, for which the background
signal of granulation is much weaker and the lower-frequency modes are
easier to detect.  Such observations will hopefully become possible as
the Stellar Oscillation Network Group \citep[SONG,][]{song} increases
its capacity.

\subsection{Comparison with parametric fits}
\label{ss:pfits}

In principle, the surface effects computed here can be used to
calibrate parametric fits available in the literature, but the
coverage in the Hertzsprung--Russell diagram (only four points) is
presently too sparse to make thorough inferences.  However, we can
compute the best-fitting parameters for these analytic models, partly
to subsequently compare with fits to real stars with similar
parameters, and partly to compare which prescriptions fit our results
better.  The latter approach is similar to the recent work by
\citet{schmitt2015}, who computed theoretical frequency shifts using
structure kernels, except that here we are using the simulation data.
\citet{sonoi2015} also compared parametric fits but limited their
comparison to the power law proposed by \citet{kbcd2008} and their own
modified Lorentzian parametrization.

Figs~\ref{f:F3}, \ref{f:G2}, \ref{f:K0} and \ref{f:K5} show the
predicted surface effects in the F3, G2, K0 and K5 models, along with
parametric fits from several sources.  Each parametric form specifies
the difference between the model frequency $\nu\st{mdl}$ and the
corrected frequency $\nu\st{cor}$ as some function of the model
frequency and sometimes mode inertia $\mathcal{I}$, normalized to the
total displacement at the photosphere.

First, there are the two fits presented by
\citet{ball2014}.  They proposed either a one-term fit,
\eq{\label{e:bg1}\nu\st{cor}-\nu\st{mdl}=a_3(\nu\st{mdl}/\nu\st{ac})^3/\mathcal{I}}%
or two-term fit
\eq{\label{e:bg2}\nu\st{cor}-\nu\st{mdl}=\left(b_{-1}(\nu\st{mdl}/\nu\st{ac})^{-1}+b_3(\nu\st{mdl}/\nu\st{ac})^3\right)/\mathcal{I}}%
where $a_3$, $b_{-1}$ and $b_3$ are coefficients that minimize the
difference between a model and whatever observations are being fit.
Here, the acoustic cut-off frequency $\nu\st{ac}$ is used purely to
scale the frequencies.

Second, there is the power-law fit suggested by \citet{kbcd2008},
\eq{\label{e:powerlaw}\nu\st{cor}-\nu\st{mdl}=p_0(\nu\st{mdl}/\nu\st{ref})^{p_1}} %
Those authors proposed that $p_1$ (originally denoted $b$) be
calibrated to the Sun and $p_0$ calibrated using the modelled and
observed large separations, based on homology arguments.  Here, we
optimize the values of both to see how they vary, and use the acoustic
cut-off as the reference frequency $\nu\st{ref}$.

Third, there is the modified Lorentzian suggested by \citet{sonoi2015},
\eq{\label{e:sonoi}\nu\st{cor}-\nu\st{mdl}=
s_0\nu\st{max}\left(1-\frac{1}{1-\bracfrac{\nu\st{mdl}}{\nu\st{max}}^{s_1}}\right)}
where the frequency of maximum oscillation power $\nu\st{max}$
is determined from the scaling relation \citep{kjeldsen1995}
\eq{\frac{\nu\st{max}}{\nu_{\mathrm{max,\odot}}}=\frac{g}{g_\odot}
\bracfrac{T\st{eff}}{T_{\mathrm{eff},\odot}}^{1/2}\text{.}}
The free parameters $s_0$ and $s_1$ correspond to the parameters
$\alpha$ and $\beta$ in the original work by \citet{sonoi2015}.

Finally, we include a scaled solar correction, computed as described
by \citet{schmitt2015}.  Our solar correction was taken as the
difference between Model S and the low-frequency BiSON data. i.e. the
grey points in Fig.~\ref{f:modelS}.  We note that including a constant
offset does not make sense, since it is clear that the low-frequency
modes are not shifted at all in our results.  We fit all of the
coefficients using non-linear least squares without weighting any of
the frequencies.

The best-fitting parameters are presented with the corresponding
models in Table~\ref{t:data}.  Though we refrain from concluding
anything quantitative from the unweighted residuals, it is still
useful to inspect the quality of the fits.  Figs~\ref{f:F3},
\ref{f:G2}, \ref{f:K0} and \ref{f:K5} show that the two-term fit by
\citet{ball2014} overall fits the simulations better than the other
corrections, notably including the F3 simulation.  For the F3
simulation, it is important to include the mode inertiae in the
fitting formula to correctly recover the distinct shape of the surface
effect.

The scaled solar surface term performs well in the cooler stars, where
it reproduces the initial increase in the surface term quite well.
However, because the solar surface term is scaled by mean density (or
the large separation) rather than the acoustic cut-off frequency, a
shrinking frequency range is covered as the temperature decreases.
The modified Lorentzian is also generally able to reproduce the
initial rise in the magnitude of the surface term as a function of
frequency.  In the F3 model, neither the scaled solar correction nor
the modified Lorentzian are able to match the surface term's more
complicated shape. For the G2 and F3 simulations, the parameter values
of the modified Lorentzian compare well with the fits by
\citet{sonoi2015} to their corresponding models A and B.

The power law clearly misses the distinct shape of the F3 simulation,
but also fails to reproduce the sharp increase and ultimate decrease
of the surface effect in the cooler stars.  Note that, to reduce the
error at high frequency, the indices of the power laws are all much
lower (around $2$) than typical values used in the correction by
\citet{kbcd2008} (usually between about $4.5$ and $5.5$), but
reasonably consistent for all four spectral types.

For the sake of comparison, we have included in Table~\ref{t:data} the
results of calibrating the surface corrections to the frequency
differences between Model S and the low-degree BiSON observations
(i.e. the grey points in Fig.~\ref{f:modelS}).  The overall magnitudes
of the surface terms are consistent with our results for the fitted
stellar models, though the result for the power law is markedly
different.  There are many potential reasons for slight differences in
the observed and modelled surface terms but the most important is that
we have only regarded the static structural effect of the surface
term.  The frequency differences between Model S and the BiSON
observations necessarily include all the physical processes that
contribute to the surface effect (see Sec.~1).

\section{Conclusion}
\label{s:conclusion}

We have used profiles from averaging 3D radiation hydrodynamic
simulations over time and space for four spectral types to compute
corrections to stellar oscillation frequencies induced by better
modelling the equilibrium structure of the near-surface layers of
solar-like oscillators: a component of the so-called \emph{surface
  effect}.  In the three cooler simulations (types G2, K0 and K5), the
surface effects are similar in shape to what is already known from
differences between solar observations and models calibrated to the
match the observed solar luminosity and radius at the meteoritic age
of the solar system.  The hotter simulation, of a star of spectral
type F3, predicts a qualitatively different surface effect.  Across
the four cases, the surface effect consistently decreases in magnitude
with decreasing effective temperature.  In other words, our results
suggest that hotter main-sequence stars have larger surface effects.

By comparing parametric fits available in the literature, we find that
the two-term fit by \citet{ball2014} is best able to reproduce the
frequency shifts, though the scaled solar term performs comparably
well in the solar-type and cooler stars.  Thus, we corroborate the
conclusion of \citet{schmitt2015}, who modelled surface effects using
structural perturbations.

Our derived frequency difference generally agree with the recent
results by \citet{sonoi2015}, who in essence performed the same
calculation for a different range of stellar parameters using
different modelling codes.  Our G2 and F3 simulations are similar to
their models A and B, and our derived frequency differences are
qualitatively and quantitatively very similar.  Their models generally
explored lower surface gravities, whereas our K0 and K5 simulations
extend the results to cooler main-sequence stars.  Our methods thus
corroborate their results where they overlap, and complement them
elsewhere.  \citet{sonoi2015} did not fit the parametrizations of
\citet{ball2014} and concluded that their modified Lorentzian is
superior to the power law of \citet{kbcd2008}.  Our results support
this conclusion but find that the combined correction of
\citet{ball2014} is even better, mostly because it incorporates the
mode inertiae.

This preliminary study exploits simulation data that was
serendipitously created for other purposes.  The obvious next step is
to compute further simulations specifically to be matched to stellar
models.  In addition, we intend to explore automatic calibration of
the models to the averaged simulation profiles, rather than matching
the surface properties and manually adjusting the mixing-length
parameter to obtain a good fit at the base of the MURaM simulations.

We close by noting that our results only treat one structural
component of the surface effect.  That is, we have only computed
frequency shifts caused by simplified modelling of the static,
equilibrium state of the near-surface layers.  These results have no
bearing on the effects caused by ignoring non-adiabatic effects on the
oscillation modes, perturbations to the turbulent pressure, or small-
or large-scale flows.  Any of the other components may induce surface
effects that vary differently between different stars and, taken
together, produce trends that differ from what we have found.
However, as is clear from the solar case, the structural effect is a
major contributor and this work thus offers insight into how the
surface effect varies across the main sequence.

\begin{acknowledgements}
  The authors acknowledge research funding by Deutsche
  Forschungsgemeinschaft (DFG) under grant SFB 963/1 ``Astrophysical
  flow instabilities and turbulence'', Projects A18 (WHB) and A16 (BB
  and RHC).  LG acknowledges support by the Center for Space Science
  at the NYU Abu Dhabi Institute under grant G1502.  We are grateful
  to Aaron Birch and Hans-G\"unter Ludwig for helpful comments on the
  results and manuscript.
\end{acknowledgements}

\bibliographystyle{aa}
\bibliography{mm}

\end{document}